\documentstyle[11pt,aaspp4]{article}

\begin{document}

\title{Detection of Dynamical Structures\\
    Using Color Gradients in Galaxies}

\author{
A. C. Quillen\altaffilmark{1}, 
Solange V. Ram\'{\i}rez\altaffilmark{2} \&
Jay A. Frogel\altaffilmark{3}}
\affil{Astronomy Department, Ohio State University, 174 W. 18th Ave., 
    Columbus, OH 43210}

\altaffiltext{1}{E-mail: quillen@payne.mps.ohio-state.edu}
\altaffiltext{2}{Visiting Astronomer, Cerro Tololo Inter-American Observatory. 
CTIO is operated by AURA, Inc.\ under contract to the National Science
Foundation.} 
\altaffiltext{3}{Visiting Associate of the Observatories, Carnegie 
Institution of Washington.}


\begin{abstract}

We describe a technique that uses radial color gradients in disk galaxies
to detect the presence of bulk non-circular motion or elliptical orbits.
In a disk galaxy with both a radial color gradient and non-circular
motion, isochromes or iso-color contours 
should follow the shape of closed stellar orbits,   
and the ellipticity of the isophotes should vary as a function
of wavelength. 
A difference in the ellipticity of isochromes and the isophotes
can be used to detect the presence of non-circular motion.
A model galaxy is constructed which demonstrates this phenomenon.
The difference between isochrome and isophote ellipticity is directly 
related to the ellipticity of the potential.  
This provides 
a new way to measure the ellipticity of the dark matter in 
the outer parts of galaxies.

As an example, we apply this technique to 
two dwarf galaxies NGC 1800 and NGC 7764.
We detect a bar in NGC 1800 which
has only previously been suggested from the HI velocity field. 
In NGC 7764 there is no color gradient along its bar so we cannot
detect non-circular motion in this region; however ellipticities observed
in a star forming
ring at the end of the bar are consistent with this ring being located
near the corotation resonance.
\end{abstract}


%

\section { INTRODUCTION }

Previous studies of observational signatures of non-circular 
or elliptical motion 
in galaxies include  Franx et al.\ (1994), Sackett et al. (1994), 
Franx and de Zeeuw (1992), Staveley-Smith et al. (1990), and 
de Zeeuw and Franx (1989).  These studies explore how 
twists in the velocity field coupled with measurements of the ellipticities 
of the isophotes (in one wavelength band) in a galaxy
can be used to constrain the ellipticity of the halo
or the axis ratios of a triaxial gravitational potential.
Many of these studies have noted  (e. g.
Franx and de Zeeuw 1989 and  Franx et al. 1994) that there should
be a difference between isophote ellipticity and orbit ellipticity,
however this difference has not been exploited by using
the information available in color maps.

Multi-band imaging studies of late-type galaxies have found that these
galaxies almost always have radial color gradients (e.g., \cite{dej94}).
Color maps often reveal 
underlying structures such as bars and rings (e.g. Buta 1990).
Although color gradients have been used to study the dynamics
of different stellar populations in elliptical galaxies
(e.g. Thomsen \& Baum 1989), little
been done using color gradients to derive dynamical information
in spiral galaxies.
In this paper we introduce how the ellipticity of the isophotes 
should vary as a function of wavelength in a disk galaxy that has both
a radial color gradient and non-circular stellar motion.
This effect, once detected, can be used to detect and study bars and other
non-axisymmetric structures such as a triaxial halo.
We illustrate with a model galaxy which has a color gradient
that when the potential is non-axisymmetric the isophotes
have different shapes than the isochromes.  We show how
the ellipticity of the potential can be derived from 
the orbit and isophote ellipticities.

To illustrate a preliminary use of this phenomenon,
we search for it in two dwarf galaxies NGC 1800 and NGC 7764.
Dwarf galaxies are a promising place to search for this effect because
they commonly harbor bars and other asymmetric structures. 
Their low dust content facilitates measuring the ellipticity
of the isophotes, and they commonly have color gradients (e.g. 
\cite{qui95}).
Our data are a preliminary part of a survey of $200$ to $300$
galaxies that will produce a library of photometrically calibrated
images of late-type galaxies from $0.4$ to $2.2 \mu$m.


\section { COLOR GRADIENTS AND NON-CIRCULAR MOTION}

In disk galaxies the velocity dispersion of the stars
is typically far smaller than the rotational
velocity so that many stars are in nearly circular orbits in a plane.
In the epicyclic approximation stars can be considered
to be oscillating at the epicyclic frequency about circular
orbits.   If the galaxy is perturbed by
an elliptical potential, then the closed
orbits are elliptical instead of circular.
The stars are then considered
to be oscillating about a family of closed elliptical orbits.
When the stellar velocity dispersion is small as is commonly true in late type galaxies, 
the appearance of the galaxy is determined by the structure of
these closed orbits.

Consider a galaxy with a stationary bar shaped perturbation to the potential.
Closed orbits in this galaxy have an elliptical shape.
Usually the velocity of a star in a closed elliptical orbit is 
slowest at the apocenter or major axis vertex of the ellipse.
This causes the galaxy mass surface density or surface brightness 
along this orbit to increase at the apocenter.
The isophotes of the galaxy are therefore determined by 
the radial light distribution and the shapes 
and velocity variation in the orbits.  
The mass continuity equation for a steady state system
can be used to predict the surface density of the galaxy from 
the azimuthally averaged stellar surface density and the structure of
the closed orbits.  In other words, by considering the volume of a small 
parcel, the density variation about the orbit can be
determined from the velocity and volume change of the parcel.
The difference between isophote and orbit ellipticities has been
noted previously  
(e. g.  Franx and de Zeeuw 1989 and  Franx et al. 1994). 

Most galaxies exhibit radial color gradients from variations in stellar
population and dust distribution.
This means that the azimuthally averaged surface intensity
is a function of wavelength.      Differential
rotation causes the stellar population to be evenly
distributed about the closed orbits, so that these orbits
are iso-population contours.
These contours should also be isochromes in a color map.
Because the closed orbits in an axisymmetric galaxy 
are circular, the isophotes and isochromes are coincident and also circular.
However, in a non-axisymmetric galaxy
the isophote shapes depend on orbit structure
as well as the azimuthally averaged surface intensity.
This implies that 
the isophotes shapes vary as a function of wavelength when there
is both non-circular motion and a color gradient.  
The ellipticities of the isochromes        
should differ from the ellipticities of the isophotes.  
Observation of this behavior can be used to detect and study
the presence of non-circular motion.

\subsection{In the case of small ellipticity}

Orbits that are approximately elliptical with a small ellipticity 
can be described by a radius
$$
R(r_0,\theta) = r_0\left(1 + {\epsilon_r(r_0)\over 2} 
\cos 2\theta\right),
$$
where $\epsilon_r$ is the ellipticity of the orbit.
On an individual orbit we can define a speed and a surface density 
$$
v(R(r_0,\theta)) = v_0(r_0)\left(1 + {\epsilon_v(r_0)\over 2} 
\cos 2\theta \right)
$$
$$
\Sigma(R(r_0,\theta)) = \Sigma_0(r_0)\left(1 + {\epsilon_\Sigma(r_0)\over 2} 
\cos 2\theta\right).
$$
We describe the density in a ring (not along the orbit) as
$$\Sigma(r,\theta)= \Sigma_0(r)(1 + {\epsilon_D(r) \over 2} \cos 2\theta)$$
so the ellipticity of the isophote is $\epsilon_D$.
Since the density is related to the velocity using the mass continuity equation, 
$\Sigma^{-1}$ is proportional to the velocity 
times the area element ${dA \over dr d\theta}$ in a segment of a thin
shell along each orbit.
For the shape of the orbits given above the area element is
$$
{dA(R(r_0,\theta))\over dr d\theta} = 
r_0\left(1 + (\epsilon_r+ r_0 \epsilon_r') 
{\cos 2 \theta \over 2}\right) 
$$
so on the orbit the density varies with ellipticity 
$$ \epsilon_\Sigma = - \epsilon_r - \epsilon_v -  r_0 \epsilon_r'.
$$
The ellipticity of the isophotes is then 
\begin {equation}
\epsilon_D =  \epsilon_r - { \epsilon_\Sigma \Sigma_0\over \Sigma'_0 r}
\end{equation}
For an exponential disk with density $\Sigma (r) \propto \exp(-r/r_D)$
and scale length $r_D$
\begin {equation}
\epsilon_D = \epsilon_r - 
{r_D \over r}(\epsilon_r + \epsilon_v + r \epsilon'_r).
\end{equation}

For example, a logarithmic potential 
$\Phi = v_c^2 \ln s$ where $s = \sqrt{x^2/q^2 + y^2}$ and $\epsilon=1-q$,
has orbits with $\epsilon_r = \epsilon$, 
$\epsilon_v = -2 \epsilon$ and $\epsilon_\Sigma= \epsilon$ and  
isophotes with ellipticity 
$\epsilon_D = \epsilon(1 + r_D/r )$ 
(as has been previously noted in Franx et al. 1994)
for an exponential disk with scale length $r_D$.

Since the isophote ellipticity $\epsilon_D$ depends on $\Sigma_0(r)$,
it also depends upon the band in which the galaxy is observed when 
there is a color gradient.
Differences between isophotes and color contours
can be used to detect the presence of non-circular motion.
The difference in ellipticity is 
a maximum and therefore easiest to measure) where 
$\Sigma_0'(r)$ is small and the surface brightness gradient 
is shallow.

Two angles are required to describe the orientation of a galaxy,
an inclination or tilt angle, and the position angle of the tilt
with respect to the major axis of the gravitational potential.
A projected ellipse appears to have a different position angle 
and ellipticity than its true position angle and ellipticity.
If the ellipticities and position angles of the isophotes are known
at more than two radii and vary as a
function of radius, then the projection angles of the galaxy can be 
uniquely determined (see equations 
3.12-3.14 in deZeeuw and Franx 1989).

\subsubsection{Probing the ellipticity of the gravitational potential}

Once the orientation of the galaxy is known it
is possible to directly probe the shape of the gravitational 
potential in the plane of the galaxy from the true
orbit and isophote ellipticities.
In the epicyclic approximation for a gravitational potential
$\Phi(r,\theta) = \Phi_0(r) + \Phi_1(r) \cos(2 \theta)$
with ${\Phi_1 \over \Phi_0}<<1$
the size of the non-axisymmetric perturbation is (Binney and Tremaine 1987) 
\begin{equation}
\Phi_1 = -{v_c(r)^2\over 2} (\epsilon_r + \epsilon_v)(1 - {\Omega_p\over \Omega})
\end{equation}
where $\epsilon_v$ can be determined from the isophotes and
isochromes using equation (2).
Here $v_c$ is the circular velocity
given by $d\Phi_0/dr = v_c^2/r$
and $\Omega=v_c/r $ is the angular rotation rate.
If the potential is rotating,
$\Omega_p$ is the pattern speed of the non-axisymmetric
component of the potential. 
When the potential is fixed $\Omega_p=0$.

Over much of a barred galaxy stars are found in orbits 
that oscillate about a single family of closed orbits.  
However near resonances there are not only many families of closed orbits, but
regions where there are no closed orbits.  The area of a bar
which is not covered by the major families of 
closed orbits is generally small (e.g. Athanassoula
1992) and so the effects of the resonances are limited.  
The complicated behavior of orbits near resonances
increases the stellar velocity dispersion so 
the stellar population will appear to be be well mixed.
This results in a smoothing of the color variations across the resonances.
Examples of this can be seen in Buta (1990) where color
maps show a smooth transition between orbits aligned perpendicular
and parallel to the bar.
Over much a bar the approximation that
all orbits are near closed orbits should be valid and so it is possible
to use the difference between isophote and orbit shapes to derive 
dynamical information.
Where this approximation fails a smooth variation in ellipticity should
be observed in the galaxy and equation (3) will give a spurious low value
for $\Phi_1$.

In section we have explored the limit of small
ellipticity.  However, even when the orbits are not
elliptical 
the speed variation along the orbit can still be determined
from variation in density and the area element in a segment along the orbits.
The orientation of velocity vectors can be found from the shapes of the
orbits and the speed is determined from the isophotes. 
If the rotation curve is known the entire galaxy
potential can in principal be completely reconstructed.

\subsection{A Model Galaxy Example}
In this subsection we demonstrate with a model galaxy 
that when the orbits are non-circular the isophotes and isochromes
have different shapes.

Using Newton's equations,
we have integrated stellar orbits in a logarithmic potential of the form
$\Phi = v_c^2 \ln(s)$
where $s=\sqrt{x^2+ y^2 (1 - \epsilon)^2}$ 
and $\epsilon$ sets the size of the non-axisymmetric perturbation.
For $\epsilon=0$ this potential gives a flat rotation curve
with circular velocity $v_c$.
Over a range of radius, closed orbits were found by minimizing the difference in positions
along the major axis before and after one revolution about the nucleus. 
A density array was constructed from the positions of the stars at
equal timesteps along each orbit 
by assuming that the density along the major axis $a$ of each orbit
was exponential $\Sigma(a) \propto \exp{(-a/h(\lambda))}$ with scale length 
$h(\lambda)$ a function of wavelength $\lambda$.  

In Figure 1 a model is displayed with color maps.
Figure 1a shows isophotes for a model with identical orbits but with
different scale lengths.   Note that the shape of the isophotes depends on
the scale length.  Figure 1b shows isophotes and color contours.  The color
contours are coincident with the closed orbits, 
and have a different ellipticity
than the isophotes.

\subsection{Underlying Assumptions and Caveats}
We have assumed that the stellar population in the galaxy is 
an evenly distributed dynamically relaxed and mixed stellar population.
One important effect not considered here is the role of 
star formation.  If stars are formed quickly compared to the orbital
timescale then the
colors observed will trace the local occurance of star formation and 
not directly provide information about the stellar dynamics.
This problem could be made less severe by studying near-infrared
images which are less sensitive to the
presence of young stars, and instead trace a more evenly distributed older 
stellar population.

The stellar velocity dispersion (both in the plane and vertical) 
if it is not small compared
to the rotational velocity will also affect both the projected surface
density and isochromes.  The phenomenon discussed here 
cannot be used to detect non-circular motion
near a galaxy bulge or in a very early type galaxy because these systems
are not planar and have a large velocity dispersion compared to their
rotational velocity.

The orbits of gas and dust when there are no strong shocks present 
should be almost identical to the closed stellar orbits so isochromes
should still trace the shapes of the closed orbits
even when a moderate amount of dust is present.
However when shocks are present in the ISM of a galaxy such
as is found in dust lanes in barred galaxies, it would be extremely
difficult to detect subtle effects due to population gradients
if the color maps are dominated by extinction.

We note that in the presence of non-circular orbits an inclined galaxy 
will have twists both in the isophotes (e.g., \cite{deZ89}) and isochromes.  
Warped galaxies also have twists in the
isophotes, however if the orbits are close to circular, then
the isophotes should
have the same ellipticity as the isochromes even if the
isophotes are twisted.  
At high inclination the thickness
of the stellar disk will also affect the ellipticities of the isophotes
so this effect can only be used for galaxies at low inclination.


\section {OBSERVATIONS}

Two dwarfs galaxies NGC 1800 and NGC 7764 were observed in the
near infrared $J,H$ and $K$ bands and in the visible $B,V$ and $R$ bands.

\subsection{Near Infrared Images}

The $J,H$ and $K$ images were obtained 
on the $1.0$m Swope Telescope at Las Campanas Observatory 
with the NICMOS 3 infrared array (\cite{per92}) which 
covered a field of $3.9\times 3.9$ arcminutes, with a spatial scale of $0.92$
arcsec/pixel.  
Individual images were taken with an exposure time of $60$
seconds in $J$, $20$ or $30$ seconds in $H$, and $15$ seconds in $K$.  
The observations are summarized in Table 1.
For these observations, sky was observed for a total integration
of about half of the total on source integration time.  Flat fields were
constructed from median filtered sky frames.  Images were aligned to the
nearest pixel and combined after a slight non-linearity correction, 
flat fielding and sky subtraction.  

The images were calibrated on the CTIO/CIT system   
using standard stars from Elias et al. (1982).
For NGC 7764 we calibrated the long exposure images observed
under non-photometric conditions by comparing
the magnitudes of 5 field stars in the long exposure images with the magnitudes
measured on short exposure images observed under photometric conditions.
From the scatter in the magnitudes of these stars and the
standards we estimate our accuracy 
of the calibration to be within $\pm 0.05$ mags.
There is good agreement between our calibrated images of NGC 1800 and
published aperture photometry by 
Hunter \& Gallagher (1985) with an aperture diameter of $23''$. 
We find that our $J$, $H$ and $K$ magnitudes are brighter than
those of above authors by $0.07$, $0.08$, $0.02$ mag at $J$, $H$, and $K$
respectively. 
The FWHM of stars in the final images are $\sim 2''\hskip-2pt .5$
which is larger than that caused by seeing
because of the shifting of individual images 
required to construct the final images.

\subsection{Optical Images}

The $B$, $V$, and $R$ images were observed under photometric conditions
on the 0.9m telescope at CTIO,
using a $1024 \times 1024$ pixel CCD with a spatial scale
of $0.40''$/pixel.  Individual images were taken with an exposure time of
$300$ seconds in $B$, $200$ seconds in $V$, and $120$ seconds in $R$.
A linear overscan bias and dark images were subtracted from the individual
images, which were then flat fielded using flat fields constructed 
from means of dome flats.
Cosmic ray events were removed using interactive median filtering.
The final images were made from averages of the individual images.
The calibration was done using standard stars from Landolt (1992).  
We estimate our accuracy to be  $\sim \pm 0.04$ mag. 
There is good agreement between our calibrated images of NGC 1800 and
published aperture photometry by
Gallagher \& Hunter (1987) with an aperture diameters of
$57''$ and $130''$.
We find that our $B$, $V$ and $R$ magnitudes are brighter than
those of above authors by $0.03$, $0.03$, $0.02$ mag at $B$, $V$, and
$R$ respectively, but consistent within the errors.

\subsection{Registered Images and Color maps}

To transform the optical and infrared images to the same pixel scale,
orientation and resolution, we used 
linear transformations which included a rotation and a scaling term as
well as a shift.  These transformations were derived from measurements
of centroids of stars in the field.
We estimate that our registration is good to within $0.3''$ across
the images based upon the standard deviation of centroids of 
the field stars of the registered images.  
The images were then smoothed so that stars had the same
FWHM in all bands.  
Registered, resampled and smoothed 
optical and near infrared images are displayed in Figure 2.


\section {APPLICATION TO NGC 1800 AND NGC 7764 }

Both NGC 1800 and NGC 7764 have blue central regions and a patchy 
amorphous appearance probably because of recent star formation.
Except for two bright spots in the images of NGC 1800
the isophotes are regular and elliptical even in the $B$ band.
The HI velocity field of NGC 1800 suggests that this galaxy 
harbors a bar (\cite{hun94}).
NGC 1800 has a disturbed H$\alpha$ morphology with
plume-like and loop-like features (Hunter etal 1994).
However the morphology of the visible images does not
resemble the morphology of the H$\alpha$ emission so that
even in the visible images primarily what is seen 
is a smooth underlying evenly distributed stellar population.
The appearance of the optical images is almost identical
to that of the infrared images.  This is consistent with
with the interpretation that primarily what is seen in all bands
is an evenly distributed dynamically relaxed stellar population.
This assumption is required to interpret a difference in
the ellipticities of the isophotes and isochromes as being
due to non-circular motion and not due to local star formation.


In Figure 3 we show $B-V$ color contours plotted with the $R$ band isophotes 
for NGC 1800 and NGC 7764.  Optical bands were chosen for measurement
of ellipticities because of their higher signal to noise.
In both galaxies there is a smooth difference in morphology between
the color maps and the isophotes.  
We checked that this difference is not a result of improper sky subtraction
by considering all sky values for both colors that are within our
estimated errors for the sky value.

NGC 1800 has color contours that are more
elliptical than the isophotes.  
If the stars were on circular orbits in
this galaxy, the color isophotes would lie on top of the isophotes, 
so the difference in the two contour maps is good
evidence that NGC 1800 does harbor a bar and we confirm the 
suggestion of Hunter et al. (1994) that NGC 1800 has a bar.
We expect that in each closed stellar orbit the density
should be largest along the bar where the velocity is lowest. 
However, if the radial gradient of the density
is large and the ellipticities of the orbits vary, then the isophotes
may be less elliptical than the isochromes (see equation 2).

In Figure 4 we plot ellipticities and position angles 
as a function of radius for both galaxies for the optical
bands where higher signal to noise permitted isophote fitting.
In NGC 1800 we note that the ellipticities of the isophotes in the
different filters
are equal at $\sim 50''$.  This would be consistent with the galaxy
having bar corotation radius at this radius.  
This radius is also suggested as the approximate
location of corotation from the HI velocity field because
the iso-velocity contours only twist inside this radius.

Using equations (3.12-3.14) in deZeeuw and Franx (1989) which
describe the projected ellipticity and position angle of an ellipse 
given its orientation and true ellipticity and position angle we find
that an inclination $\sim 81^\circ$ (close to edge on) 
and position angle $\phi \sim 70^\circ$ come closest to predicting
the range of ellipticities and position angles observed as a function
of radius in $B$ band.  
In the principal plane of the galaxy this position angle, $\phi$, is defined 
to be the angle between the line of sight
(projected onto this plane) and the major
axis of the gravitational potential (as in deZeeuw and Franx 1989).
The inclination is set by the point where the ellipticities cross
at $50''$, and the position angle was chosen so that the 
the position angle and ellipticity observed at every radius
were consistent with an ellipse viewed at these projection angles.
Using these projection angles we estimate true ellipticities for the orbit
$\epsilon_r \sim 0.35$ at $30''$,  and $\sim -0.1$ at 60'',
and for the isophotes at $B$ band
$\epsilon_D \sim 0.2$ at $30''$,  and $\sim -0.15$ at 60''.
For an exponential scale length at $B$ band of $r_D \sim 10''$ which we
have measured from the major axis of the $B$ band image, 
using equation (3) we estimate that $\Phi_1$ ranges from 
$0.45 v_c^2 (1 - {\Omega_p \over \Omega})$ at $30''$
to 
$0.3  v_c^2 (1 - {\Omega_p \over \Omega})$ at $60''$.
That the bar potential varies smoothly but is more elliptical 
within the bar than without is consistent with estimates of the gravitational
potential in barred galaxies (e.g. Quillen et al 1994).
These estimates are meant to be primarily illustrative, since they are crude.
A better data set on a galaxy that is closer to face on would warrant a 
more detailed study.

In NGC 7764's  color map we can see a blue star forming ring that
is outside the end of the bar.  The spotty appearance implies that
we are seeing local star formation in the ring, so that the light
may not trace a conserved quantity 
in an orbit, though we do expect that the stars have formed along
closed orbits.  No difference between in the isochromes 
and the isophotes
were found within the bar because there is little color gradient
in this region (see Figure 3).
We can however, use the ellipticities of the isophotes to find the end
of the bar.  The ellipticities are about the same at a radius of $\sim 40''$.
This is consistent with this ring being at the corotation radius (see
Figure 4).

\section {SUMMARY AND DISCUSSION}

In this paper we have introduced how color gradients can be used
to detect the presence of non-axisymmetric dynamical motion  
such as a bar or triaxial halo.
In a galaxy with both a color gradient and non-circular motion,
the shapes of closed orbits should follow the isochromes,
and the ellipticity of the isophotes should vary as a function of
wavelength. The difference between isophote and orbit ellipticity
should be largest where the surface brightness gradient is shallowest.
From the position angles and ellipticities of the isophotes
at different radii the orientation of the galaxy can be 
determined.  Once the orientation of the galaxy is known, the strength
of the non-axisymmetric component of the gravitational potential
can be directly measured.

To illustrate this phenomenon we have constructed a simple model
with a flat rotation curve.  This model shows that when the orbits
are elliptical the isophote shapes depend on the exponential
scale length of the disk and that the isophotes 
are not coincident with the isochromes.

We search for this phenomenon in two dwarf galaxies.  In NGC 1800
isochromes have a different ellipticity than the isophotes
implying that there is a bar in this galaxy which has 
previously only been suggested from the HI velocity field.
The ellipticities of the isophotes in different bands are equal at 
a radius of 50'' which suggests that the bar ends at this
radius. Using the observed ellipticities and position angles at different radii we
have estimated the orientation of the galaxy.  From the true ellipticities 
we find that the strength of the non-axisymmetric component of the 
potential near the end of the bar 
decreases as a function of radius as expected in a barred galaxy.

In NGC 7764 there is little color gradient along its bar so that we
do not detect the presence of non-circular motion within the bar.
At the end of the bar there is a blue ring undergoing star formation where
the ellipticities of the isophotes in different bands are equal.
This is consistent with this ring being near the corotation radius.

The recent availability of deep multi-band images of galaxies 
suggests a number of settings
for an investigation of the phenomenon introduced in this paper.
Since exponential disk radii  differ as a function 
of wavelength by a few percent (\cite{dej94}), it should be possible to detect
non-circular motion in the outer parts of disk galaxies where dark
matter is dominant.
The largest color gradient can be achieved by comparing optical and
infrared images, however using only infrared images may be desirable
because the images should be less affected by recent star formation.
Rix and Zaritsky (1994) have recently 
discovered that most galaxies are lopsided.
We suggest that color gradients can be used to constrain the lifetime of
such asymmetries.
Comparing color contours to the shape of closed orbits integrated in
the gravitational potential of a barred galaxy (e.g. as in Quillen et al 1994)
could constrain formation mechanisms for the bar.

In elliptical galaxies isophote ellipticities and orientation angle
have been observed to depend upon wavelength (e.g. Peletier 1989).  
By summing over orbits of different morphology the shapes of
elliptical galaxies could also be measured.

Since many rings of ring galaxies are located where the dark matter 
contribution is significant, 
this technique can be used to compare the morphology
of dark matter with that of non-dark matter.
In the ring of a ring galaxy which has an inner bar, 
the gravitational potential is not circular because of the bar.   
Since these galaxies typically have color gradients
the effect described in this paper should be detectable in such a situation.  
The ellipticity of the potential can be measured
as a function of radius and directly compared to the ellipticity
of the potential predicted from the light distribution.

\acknowledgments

We are grateful to Roberto Aviles for obtaining the $B,V$ and $R$ 
images for us.
We acknowledge helpful discussions and correspondence 
with G. Rivlis, R. Q. Rivlis, R. J. Buta, B. Elmegreen, H-W. Rix,
A. Gould and D. Zaritsky. 
The OSU galaxy survey is being supported in part by NSF grant AST 92-17716.
The Las Campanas Observatory IR camera was built with partial
funding from NSF grant AST-9008937 to S. E. Persson.
J.A.F's research is supported in part by NSF grant AST 92-18281.
A.C.Q. acknowledges the support of a Columbus fellowship.
J. A. F.  thanks the Physics Department of the University of Durham and PPARC 
for support via a Senior Visiting Research Fellowship.
We thank the referee, J. S. Gallagher, for helpful comments and
suggestions which improved the paper.

\clearpage

\begin{deluxetable}{crrrrrrrrrrr}
\footnotesize
\tablecaption{Summary of Observations. \label{tbl-1}}
\tablewidth{326pt}
\tablehead{
\colhead{Date} & \colhead{Galaxy}   & \colhead{photo?\tablenotemark{a}} & \colhead{$J$\tablenotemark{b}}   & \colhead{$H^b$} & \colhead{$K^b$}   & \colhead{$B^b$} & \colhead{$V^b$}   & \colhead{$R^b$} 
} 
\startdata
1992 Nov 2 & NGC 1800 & yes & 20 &    & 18 &    &    &     \cr
1992 Nov 3 & NGC 1800 & yes &    & 20 & 24 &    &    &     \cr
1994 Nov 1 & NGC 1800 & yes &    &    &    & 15 & 11 & 6   \cr
1992 Nov 5 & NGC 7764 & no  & 20 & 20 & 48 &    &    &     \cr 
1993 Oct 4 & NGC 7764 & yes & 5  & 2  & 2  &    &    &     \cr 
1994 Nov 1 & NGC 7764 & yes &    &    &    & 12 & 6  & 6   \cr 
 
\enddata

\tablenotetext{a}{A yes means the night was photometric}
\tablenotetext{b}{On source exposure times given in minutes}

\end{deluxetable}

%
%

\clearpage
\figcaption[fig1.eps]{
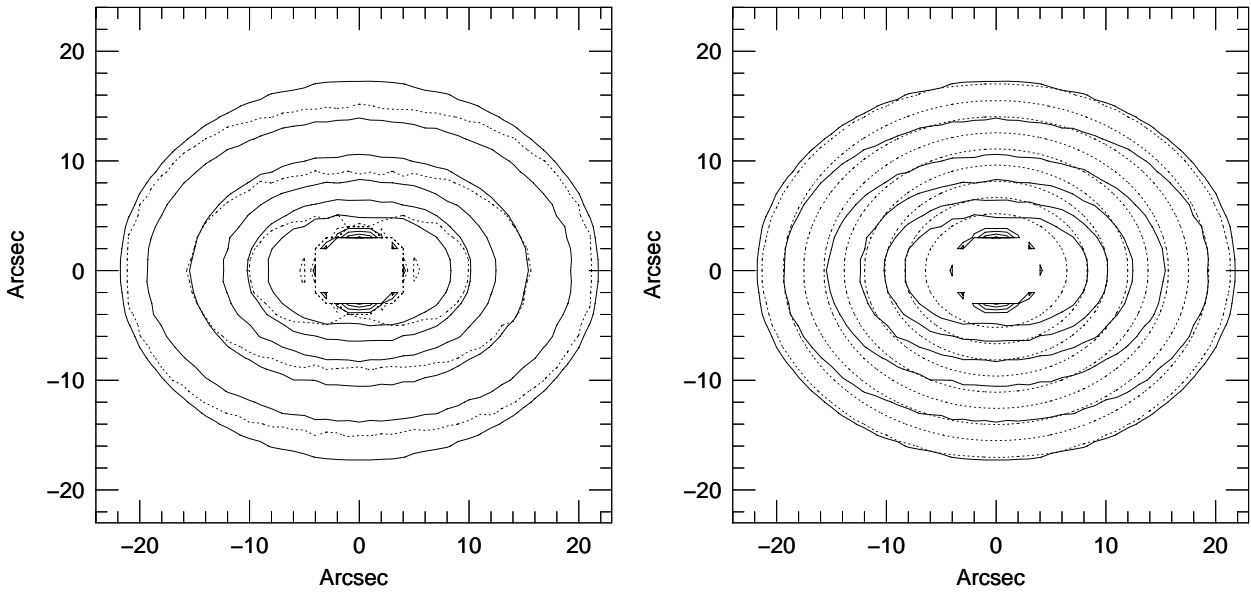
a) Isophotes for the model galaxy described in section 2.1.  
Shown are isophotes for exponential disk scale lengths $h=1$ (solid lines) 
and $h=2$ (dotted lines).
For this model $\epsilon=0.2$ and $v_c=1$.
Note that the isophote shapes depend on the exponential disk scale length.
b) Isophotes (solid lines) and isochromes (dotted lines) for the model galaxy. 
Isophotes are shown for exponential scale length $h=1$.
Note that the isophotes differ from the isochromes.
\label{fig1}}

\figcaption[fig2.eps]{
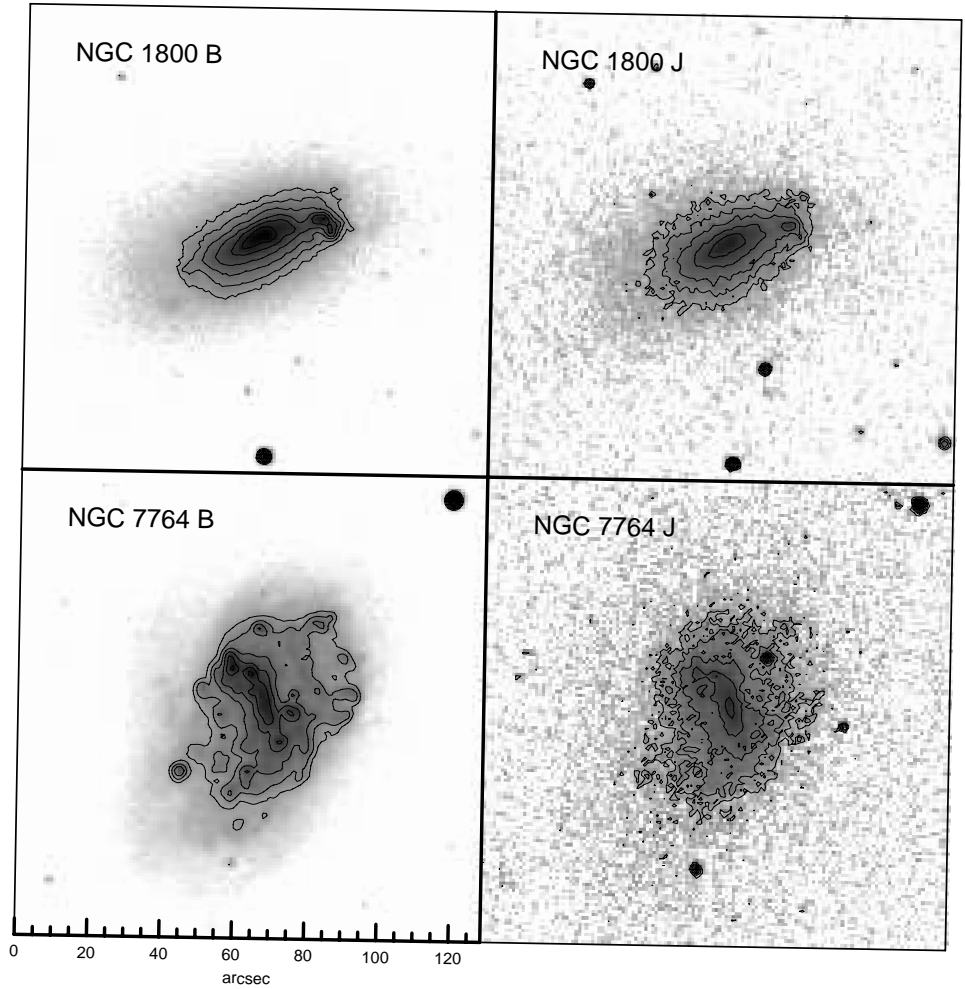
Gray scale images of the two dwarf galaxies NGC 1800 and NGC 7764 in $B$
and $J$ bands.  Contours are 0.5 magnitude apart with brightest contours at
15.5 mag in $J$ band and 13.2 mag in $B$ band.
\label{fig2}}

\figcaption[fig3a.eps]{
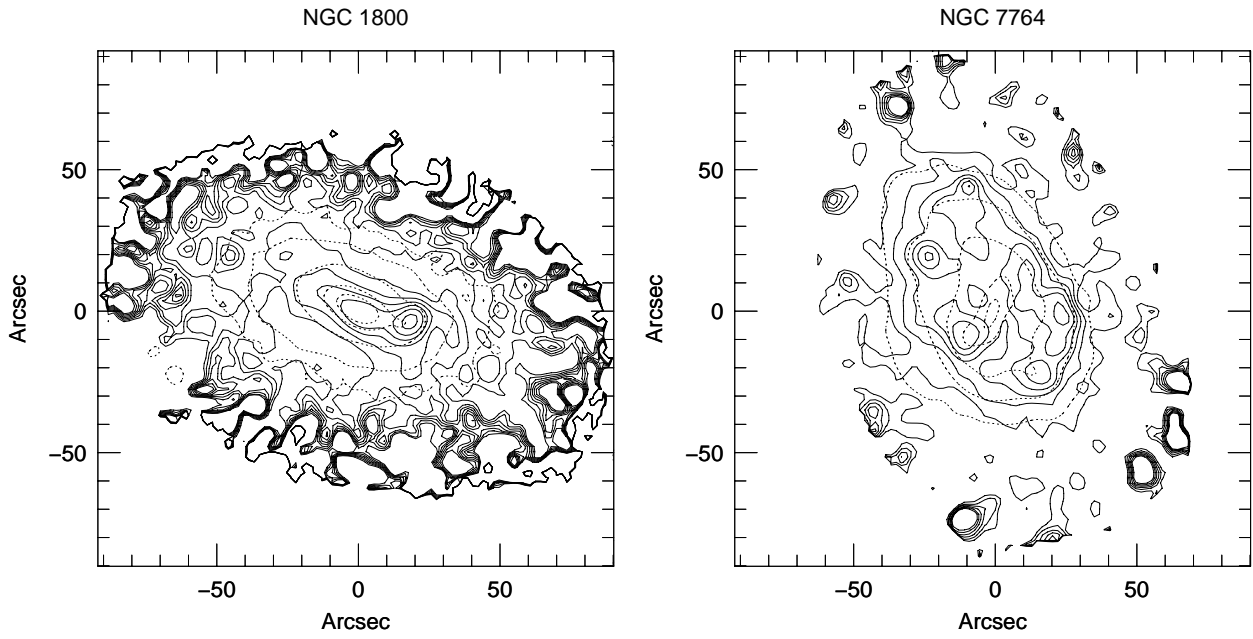
Contour plots show the $R$ band isophotes (dotted contours) and $B-V$  color
maps (solid contours) for NGC 1800 and NGC 7764. Note that
the isochromes do not like on top of the isophotes.
\label{fig3}}

\figcaption[fig4a.eps]{
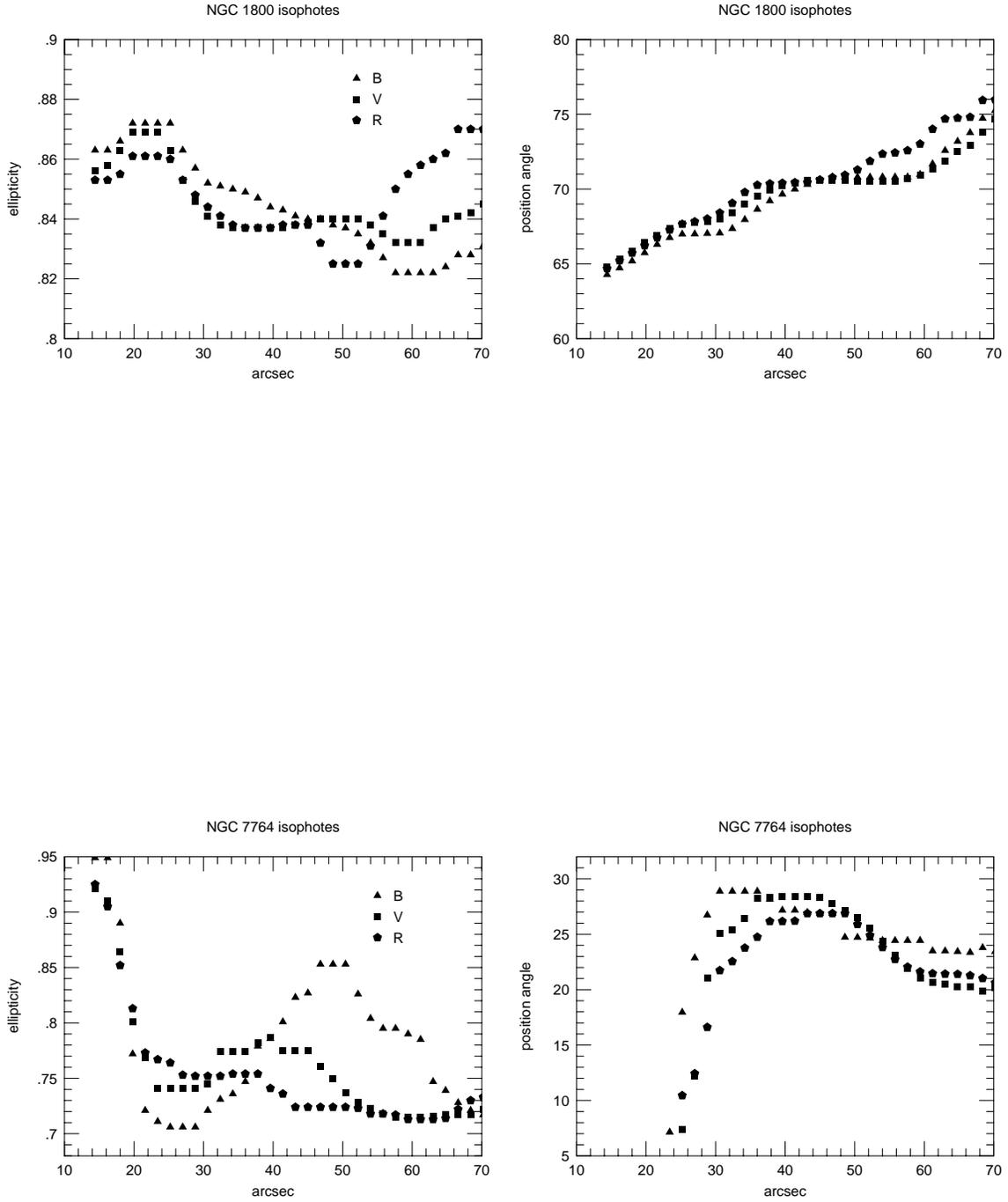
Plotted are the ellipticities of the isophotes and their position
angles (from north) as a function of radius for NGC 1800 and NGC 7764.
Note that the ellipticities cross at around 50'' for NGC 1800
and at around 46'' for NGC 7764.  
This is consistent with bars ending at these radii.
Contour fitting was done with the software Vista on images with stars removed.
\label{fig4}}

\end{document}